\documentclass{emulateapj}

\shorttitle{Photosphere deathline and GRB 110721A}
\shortauthors{Zhang et al.}

\newcommand{\beq}{\begin{equation}}
\newcommand{\eeq}{\end{equation}}
\newcommand{\ba}{\begin{array}}
\newcommand{\ea}{\end{array}}

\newcommand{\ee}{\epsilon_{e,0}}

\def \etal{{\it et al.~}}
\def\be{\begin{equation}}
\def\ee{\end{equation}}
 \def\lesssim{\raisebox{-0.3ex}{\mbox{$\stackrel{<}{_\sim} \,$}}}
 \def\gtrsim{\raisebox{-0.3ex}{\mbox{$\stackrel{>}{_\sim} \,$}}}

\renewcommand\refname{References and Notes}

\begin{document}
\title{GRB 110721A: photosphere ``death line'' \\
and the physical origin of the GRB ``Band'' function}

\author{Bing Zhang\altaffilmark{1,2}, Rui-Jing Lu\altaffilmark{1},
En-Wei Liang\altaffilmark{1}, Xue-Feng Wu\altaffilmark{3}}
\altaffiltext{1}{Department of Physics and GXU-NAOC Center for
Astrophysics and Space Sciences, Guangxi University, Nanning
530004, China}
\altaffiltext{2}{On leave from Department of Physics and Astronomy,
University of Nevada, Las Vegas, NV 89154, USA}
\altaffiltext{3}{Purple Mountain Observatory, Chinese Academy of
Sciences, Nanjing 210008, China}

\begin{abstract}
The prompt emission spectra of gamma-ray bursts (GRBs) usually
have a dominant component that is well described by a phenomenological
``Band'' function. The physical origin of this spectral component
is debated. Although the traditional interpretation is
synchrotron radiation of non-thermal electrons accelerated in
internal shocks or magnetic dissipation regions, a growing trend
in the community is to interpret this component as modified thermal
emission from
a dissipative photosphere of a GRB fireball. We analyze the
time dependent spectrum of GRB 110721A detected by {\em Fermi}
GBM and LAT, and pay special attention to the rapid evolution
of the peak energy $E_p$. We define a ``death line'' of
thermally-dominated dissipative photospheric emission in the
$E_p - L$ plane, and show that $E_p$ of GRB 110721A at the earliest
epoch has a very high $E_p \sim 15$ MeV that is beyond the ``death
line''. Together with the finding
that an additional ``shoulder'' component exists in this burst
that is consistent with a photospheric origin, we suggest that
at least for some bursts, the ``Band'' component is not
from a dissipative photosphere, but must invoke
a non-thermal origin (e.g. synchrotron or inverse Compton) in
the optically thin region of a GRB outflow. We also suggest
that the rapid ``hard-to-soft'' spectral evolution is consistent
with the quick discharge of magnetic energy in a
magnetically-dominated outflow in the optically thin region.
\end{abstract}


\slugcomment{}

\section{Introduction}
\label{sec:intro}

The prompt emission spectrum of a gamma-ray burst (GRB) is usually well
described by a phenomenological function known as the ``Band''
function (Band et al. 1993). This model, which is essentially a
broken power law function with a smooth (exponential) transition,
was traditionally invoked to model spectra of GRBs detected by
BATSE on board the Compton Gamma-Ray Observatory. The function
is found successful to describe most GRB spectra detected by
later missions as long as the spectral band is wide enough
(e.g. Abdo et al. 2009; Zhang et al. 2011).

The physical origin of this phenomenological Band function is
not identified. The traditional model is synchrotron emission
of non-thermal electrons in an optically thin region, e.g.
internal shocks or internal magnetic dissipation regions
(M\'esz\'aros et al. 1994; Tavani 1996; Daigne \& Mochkovitch
1998; Lloyd \& Petrosian 2000; Bosnjak et al. 2009;
Zhang \& Yan 2011). Alternatively, a matter-dominated
outflow (fireball) can have a bright photosphere (Pacz\'ynski 1986;
Goodman 1986; M\'esz\'aros \& Rees 2000;  M\'esz\'aros et al.
2002), which may be enhanced by kinetic or magnetic dissipation
processes near the photosphere (Thompson 1994; Rees \& M\'esz\'aros
2005; Pe'er et al. 2006; Giannios 2008; Beloborodov 2010;
Lazzati \& Begelman 2010; Ioka 2010).
It has been argued that due to geometrical and/or
physical broadening, this quasi-thermal component may be modified
to mimic a Band function (e.g. Beloborodov 2010; Lazzati \&
Begelman 2010; Pe'er \& Ryde 2011; Lundman et al. 2012).
The scalings of a non-dissipative photosphere model are argued
to be able to interpret various empirical correlations (Fan
et al. 2012).

Within the framework of the standard
fireball-shock model, a GRB prompt emission spectrum is expected
to be the superposition of a quasi-thermal photosphere emission
component and a non-thermal component in the optically-thin
internal shock region (M\'esz\'aros \& Rees 2000; Zhang \& M\'esz\'aros
2002; Toma et al. 2011; Pe'er et al. 2012).
Such a superposition effect has been claimed in the BATSE data
archive (e.g. Ryde 2005; Ryde \& Pe'er 2009), and was confirmed
more robustly recently with the Fermi data (e.g. Ryde et al. 2010;
Zhang et al. 2011; Guiriec et al. 2011; Axelsson et al. 2012).
On the other hand, most GRB spectra (e.g. GRB 080916C) are still
well described by one single Band component (Abdo et al. 2009;
Zhang et al. 2011). This sharpens the debate regarding the origin
of the Band function. For GRB 080916C, the non-detection of
a thermal component led to the suggestion of a
Poynting-flux-dominated outflow (Zhang \& Pe'er 2009; see also
Daigne \& Mochkovitch 2002; Zhang \& M\'esz\'aros 2002).
Alternatively, some authors attempted to interpret the entire
Band function as emission from a dissipative photosphere
(e.g. Beloborodov 2010; Vurm et al. 2011; Giannios 2008;
Ioka 2010). These two interpretations invoke distinct assumptions
regarding the composition of the GRB jets. Finding observational
clues to differentiate between them is
therefore essential to unveil the physics of GRB central engine,
jet composition and energy dissipation mechanisms, which are
poorly constrained (e.g. Zhang 2011).

Here we show that the time-resolved spectral information of GRB
110721A holds the key to address this open question.

\section{``Death line'' of GRB baryonic photosphere emission in the
$E_p-L$ plane}

For a hot fireball with total wind luminosity $L_w$ launched from
an initial fireball radius $R_0$, the initial temperature is
\begin{equation}
T_0 \simeq (L_w/4\pi R_0^2 c a)^{1/4} \simeq 1.4 \times 10^{10}
~{\rm K}~ L_{w,52}^{1/4} R_{0,7}^{-1/2},
\label{T0}
\end{equation}
where $c$ is speed of
light, and $a=7.56\times 10^{-15}~{\rm erg~cm^{-3}~K^{-4}}$ is
the Stefan-Boltzmann energy density constant. The observed photosphere
temperature $T_{ph}$ can be as high as $T_0$ if the fireball is clean
enough so that the photosphere radius $R_{ph}$ does not exceed the
fireball coasting radius $R_c$, but is lower than $T_0$ for fireballs
with a heavier baryon loading when $R_{ph} > R_c$. More
specifically, one has (M\'esz\'aros \& Rees 2000)
 \begin{equation}
 \frac{T_{ph}}{T_0} = \left\{
 \begin{array}{ll}
  \left(\frac{R_{ph}}{R_c}\right)^{-2/3} = \left(
\frac{\eta}{\eta_*}\right)^{8/3}, & \eta < \eta_*, R_{ph} > R_c, \\
  1, & \eta > \eta_*, R_{ph} < R_c,
 \end{array}
 \right.
\label{Tph}
\end{equation}
where $\eta=L_w/ \dot Mc^2$ is the dimensionless entropy of the
fireball, and
\begin{equation}
 \eta_* = \left(\frac{L_w \sigma_{_T}}{4\pi m_p c^3 R_0}\right)^{1/4}
 \simeq 1.04\times 10^3 \left(\frac{L_{w,52}}{R_{0,7}}\right)^{1/4}
\label{eta*}
\end{equation}
is the critical value of $\eta$.
The photosphere luminosity is $L_{ph} \simeq \pi (1/\Gamma)^2 R^2 \sigma
T^4 \propto (R/\Gamma)^2 T^4 \propto R^2 \Gamma^2 {T'}^4$. For
$\eta > \eta_*$, since $\Gamma \propto R$ and $T' \propto R^{-1}$,
one has $L_{ph} \propto R^0 \sim$ const. For $\eta < \eta_*$, since
$\Gamma \propto R^0$, $T \propto R^{-2/3}$, $L_{ph} \propto R^2
R^{-8/3} \propto R^{-2/3}$. So the photosphere luminosity is
\begin{equation}
 \frac{L_{ph}}{L_w}  = \left\{
 \begin{array}{ll}
  \left(\frac{R_{ph}}{R_c}\right)^{-2/3} = \left(
\frac{\eta}{\eta_*}\right)^{8/3}, & \eta < \eta_*, R_{ph} > R_c, \\
  1, & \eta > \eta_*, R_{ph} < R_c,
 \end{array}
 \right.
\end{equation}

If a GRB spectrum is dominated by the photosphere emission, for a
certain observed isotropic $\gamma$-ray
luminosity $L=L_{ph}$, the spectral peak energy
$E_p$ should not exceed $\zeta k T_0$, where $\zeta$ is a factor
to denote the $\nu F_\nu$ peak of the photosphere spectrum.
Therefore a baryonic photosphere emission has a ``death line''
defined by
\begin{equation}
 E_p \leq \zeta k T_0 \simeq 1.2 ~{\rm MeV}~\zeta L_{52}^{1/4}
R_{0,7}^{-1/2}.
\label{deathline}
\end{equation}
The factor $\zeta$ is subject to the shape of the spectrum.
For a strict blackbody, $\zeta \sim 3.92$. For a relativistic
outflow, the shape of blackbody is modified to the form
(see also Li \& Sari 2008 and references therein)
\begin{equation}
 F_\nu \propto \frac{\nu^2}{c^2} kT \int_{h\nu/kT}^\infty
\frac{dx}{e^x-1}.
\end{equation}
The $\nu F_\nu$ spectrum has a maximum value at
\begin{equation}
\zeta \sim 2.82.
\label{zeta}
\end{equation}

The death line (\ref{deathline}) is derived for a non-dissipative
photosphere. For a thermally-dominated jet, dissipation via
internal shocks or neutron collisional heating near the photosphere
would compensate adiabatic cooling, so that the photosphere
temperature would be maintained close to but never exceed the
maximum temperature (Beloborodov 2012), so eq.(\ref{deathline}) applies
to a broad categories of dissipative photosphere models as
well. A possible exception would be the dissipative photosphere model
that invokes continued magnetic dissipation at extended radii
above the photosphere (Drenkhahn \& Spruit 2002), which allows
$E_p$ to be above the death line (\ref{deathline}) if the bulk Lorentz
factor is large enough and dissipation at high optical depth
is suppressed (Giannios 2012). However, in order to thermalize
the jet, the required Lorentz factor is very low (Vurm et al. 2012),
making it essentially impossible to exceed the death line. Also
this model may not interpret the GRB 110721A phenomenology, including
the existence of the shoulder thermal component and the rapid
spectral softening during the rising phase of bolometric
luminosity (see Sect. 3).

\section{GRB 110721A}

GRB 110721A was jointly detected by the Gamma-Ray Burst Monitor
(GBM; Meegan et al. 2009) and the Large Area Telescope (LAT;
Atwood et al. 2009) onboard the Fermi Gamma-Ray Telescope
(Axelsson et al. 2012 and references therein). A candidate
optical counterpart was reported (Greiner
et al. 2012). Assuming that the association is real,
Berger (2012) suggested two possible redshifts
$z=0.382$ or $z=3.512$, with the former one preferred.

The time-dependent spectral evolution of GRB 110721A was
reported by Axelsson et al. (2012). Two noticeable features of
the burst are: 1. A thermal component is identified to be
superposed on the Band component in both the time integrated
spectrum and time-resolved spectra. The temperature of this
``shoulder'' thermal component evolves with time as a broken
power law, which is consistent with the expectations of the
photosphere model (Ryde \& Pe'er 2009). 2. The Band component
displays an extremely rapid spectral evolution. The $E_p$ at
the earliest epoch reaches a record-breaking value $\sim 15$ MeV.

\begin{figure}
\includegraphics[angle=0,scale=0.9]{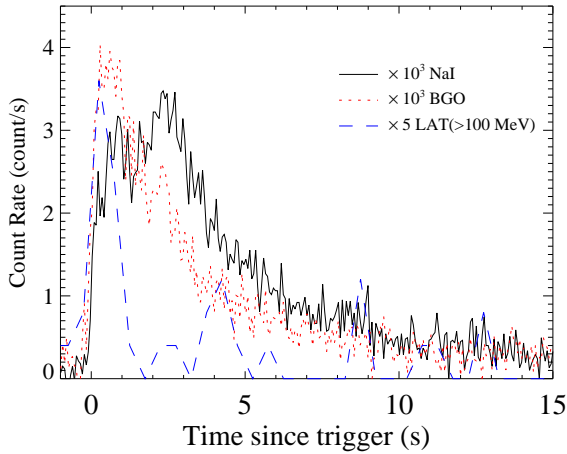}
\caption{The light curves of GRB 110721A in different energy bands:
solid: NaI; short dash: BGO; long dash: LAT above 100 MeV.} \label{LC}
\end{figure}

We have independently processed the Fermi data of GRB 110721A.
We performed a joint spectral analysis using the data from the NaI
6, 7 and BGO 1 detectors on GBM, as well as the LAT
data\footnote{http://fermi.gsfc.nasa.gov/cgi-bin/ssc /LAT/LATDataQuery.cgi}.
For GBM, we used the TTE event data containing individual
photons with time and energy tags. Background rates are estimated
by fitting the light curve before and after the burst using a
one-order background polynomial model. We pretreated the LAT data using
the LAT ScienceTools-v9r27p1 package and the P7TRANSIENT$\_$V6 response
function (detailed informatin for the LAT GRB Analysis are available
in the NASA {\em Fermi} web
site\footnote{http://fermi.gsfc.nasa.gov/ssc/data/analysis/scitools /lat$\_$grb$\_$analysis.html}).
In the diffuse response calculation, a three-component model is used:
GRB 100721A with a PowerLaw2 spectrum, the Galactic diffuse model of
gal$\_$2yearp7v6$\_$v0.fits, and the extragalactic diffuse power law
model. We then extracted the background-subtracted light curves from
the GBM and LAT and demonstrated them in Figure \ref{LC}. It is
obvious that higher energy photons arrive earlier than the lower
energy photons.

\begin{figure}
\includegraphics[angle=0,scale=0.5]{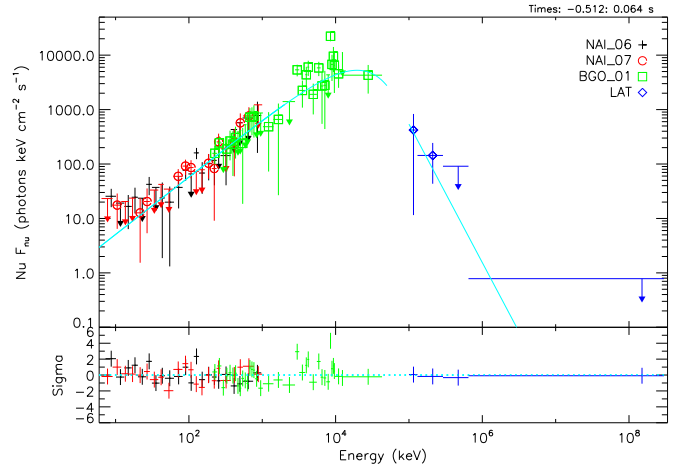}
\caption{Data and Band function model curve of GRB 110721A in the
time interval of [-0.512, 0.064]s. The solid line is the
best Band function fit with parameters
$\alpha=-0.944\pm0.046 $, $\beta=-4.543\pm0.887$, and peak energy
$E_{p}=(1.956\pm0.422)\times10^{4}$ keV, which are consistent
with Axelsson et al. (2012).} \label{spectrum}
\end{figure}

\begin{figure}
\includegraphics[angle=0,scale=0.8]{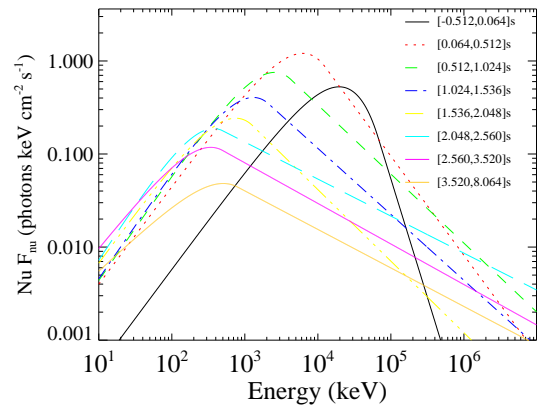}
\caption{The best-fit $\nu F_{\nu}$ model spectra for the time
resolved data in different time bins (as marked in the legend).}
\label{models}
\end{figure}

In order to better understand the cause of such a behavior, we carry
out a detailed time-dependent spectral analysis. We adjust the size
of the time bins so that each time bin contains enough photons to
perform a statistically significant spectral analysis. We applied
the software package RMFIT (version 3.3pr7) to carry out the analyses.
We confirm the conclusion of Axelsson et al. (2012) that adding a
shoulder thermal component can improve the fits to the data
significantly. Since the evolution of the thermal component has been
presented in Axelsson et al. (2012), in this paper we focus on the
Band component only in order to study its physical origin. We find
that the Band model usually gives a reasonable fit to the data, with
the reduced $\chi^{2}$ in the range $\sim (0.9-1.1)$.
Figure \ref{spectrum} gives an example of the Band model fitting
the data in the time interval of [-0.512, 0.064]s. This earliest
epoch indeed shows an extremely high $E_p$ value $\sim 19.6 \pm 4.5$
MeV, which is consistent with $E_p = 15 \pm 1.7$ MeV reported by
Axelsson et al. (2012). The larger error in our fit may be because
the Fermi team has included the extra LLE (LAT Low Energy) data,
which is currently unavailble to the public. We also present the
best fit $\nu F_\nu$ model curves in different time intervals
in Fig.\ref{models}.

\begin{figure}
\includegraphics[angle=0,scale=0.6]{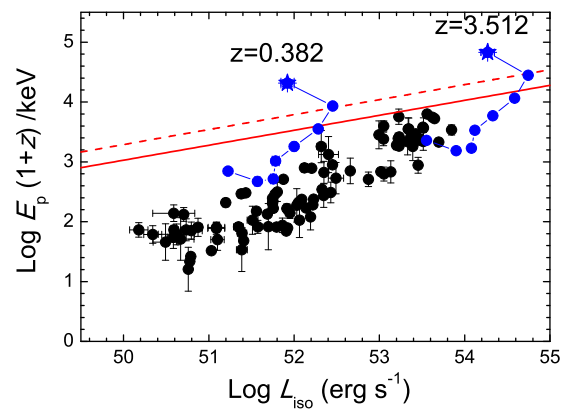}
\caption{The rest-frame peak energy $E_p (1+z)$ is plotted against
the observed isotropic $\gamma$-ray luminosity $L$ for the time
resolved spectra of Fermi GRBs with redshift measurements. The
black data points are from Fig. 9 of Lu et al. (2012), but only
the spectra during the rising phase of GRB pulses are taken
(see text for explanation). The blue points are the time-resolved
spectra of GRB 110721A for the two candidate redshifts. The two
stars are for the first epoch, which are both well beyond the
death lines. Two death lines are plotted, which
correspond to $R_0 = 10^7$ cm (solid) and $3\times 10^6$ cm
(dashed), respectively.}
\label{fig:deathline}
\end{figure}

Figure \ref{fig:deathline} displays the rest-frame $E_p - L$
plot for time-resolved spectra of Fermi GRBs with well measured
redshifts. The data are a sub-sample of Fig.9 of Lu et al. (2012),
for which only the parameters during the rising phase of GRB
pulses are adopted. This is because the decaying phase may be
controlled by the high-latitude curvature effect, which does
not directly reveal radiation physics.
Two ``death lines'' (eq.[\ref{deathline}]) are drawn, which
correspond to $R_0 \sim 10^7$ cm (solid, typical value) and
$R_0 \sim 3 \times 10^6$ cm (dashed, an extreme value to allow
highest death line possible). GRB 110721A at the earliest epoch
[-0.512. 0.064]s is plotted for two candidate redshifts.
It is clearly seen from the figure that for both redshifts, the
points are way above the death lines. This rules out
a wide range of dissipative photosphere models at least for
this earliest epoch.

One may make one step further. Axelsson et al. (2012) showed that
$E_p$ evolution of the Band component can be fit as a power-law
decay with time, suggesting that this component might have a same
physical origin during the burst. Figure \ref{models} displays
evolution of the spectral shape of the Band component, which shows
a similar $\alpha$ value with a gradually shallowing $\beta$ value.
This suggests that the Band component should share the same
physical origin in different epochs: it is not from
the photosphere, but is likely from an optically thin
region where non-thermal particles are accelerated.

\section{Summary and discussion}
\label{sec:summary}

We have shown that the Band function component of GRB 110721A is
beyond the ``death line'' of thermally-dominated dissipative
photosphere models in the $E_p - L$ plane. Together with the
fact that an additional shoulder thermal component is consistent
with the photosphere model, we reach the conclusion that the
so-called ``Band'' component is {\em not} of a photospheric
origin at least for this burst, and is formed via non-thermal
dissipation processes in the optically thin regions. Veres et
al. (2012) interpreted this emission as synchrotron emission
from a magnetically-dominated jet.

Such a finding has profound implications in understanding the
origin of other Band function spectra of GRBs. Zhang et al. (2011)
identified three elemental spectral components through a detailed
time-resolved spectral analysis of 17 GRBs co-detected with Fermi
GBM and LAT. They found that there are two types of GRBs.
GRB 080916C's spectra remain the ``Band'' shape even though the
time interval for the spectral analysis progressively reduces,
reaching $\sim 1$ s in the rest frame. GRB 090902B, on the other
hand, showed a clear ``narrowing'' feature as time bin reduces,
and spectrum is much narrower at $\sim$ rest-frame 1 s.
The time integrated spectrum of this burst is also narrower than
other Band GRBs. This burst's ``Band'' component can be indeed
de-composed as superposed photosphere emission (Ryde et al. 2010;
Zhang et al. 2011; Pe'er et al. 2012; Mizuta et al. 2011).
However, such bursts are not common.
Most bursts are similar to GRB 080916C.

The spectral parameters of GRB 110721A are similar to those of
GRB 080916C and many other GRBs. The conclusion that the Band
component of GRB 110721A originates from the optically thin
region also supports the suggestion that most Band components
are non-thermal (synchrotron or SSC) emission in the optically
thin region. Available data seem to suggest the following
unified picture: The GRB central engine may have a range of
magnetization parameter $\sigma_0$. 1. For low $\sigma_0$ bursts,
dissipation can occur at small radii, so that a bright photospheric
emission component (such as the case of GRB 090902B) is detected.
Such bursts are usually accompanied by a high energy component
due to upscattering of the thermal photons (and probably also
synchrotron self-Compton, Pe'er et al. 2012). 2. For intermediate
$\sigma_0$ bursts, the photosphere component is weaker but still
detectable. Examples of this category include GRB 110721A
(Axelsson et al. 2012) and GRB 100724B (Guiriec
et al. 2011). The Band component of these bursts are formed
in the large radii via internal shocks (IS) or internal
collision-induced magnetic reconnection and turbulence (ICMART).
3. Finally, if $\sigma_0$ is large enough, both the photosphere
and IS components are suppressed. The Band component forms at
even larger radii via the ICMART process (Zhang \& Yan 2011).

Finally, very rapid hard-to-soft $E_p$ evolution observed in GRB
110721A (Axelsson et al. 2012 and this work) and many other GRBs
(Lu et al. 2010, 2012) challenges existing models. Such a rapid
evolution is not expected in the internal shock model. For
the photosphere model, some moderate hard-to-soft evolution may
be expected due to the intial growth of optical depth in a fireball
(W. Deng \& B. Zhang 2012, in preparation), but never an extreme
evolution like GRB 110721A. A possible interpretion may be made
within the framework of the ICMART model. According
to this model (Zhang \& Yan 2011), $\sigma$ in the emission
region rapidly decreases during each ICMART event, since the
magnetic energy is continuously dissipated. If a good fraction
of local magnetic dissipation energy is deposited to electrons,
the typical electron Lorentz factor $\gamma_e$ would have
a $\sigma$-dependence, so that $E_p$ decreases with
time as $\sigma$ reduces. Also relativistic
turbulent reconnection would lead to locally Doppler-boosted
mini-jets, whose Lorentz factors would also depend on the local $\sigma$
value (Zhang \& Zhang 2012). The time dependent Doppler boosts for
mini-jets  would enhance the hard-to-soft $E_p$ evolution.

\acknowledgments
We thank the referee for helpful comments,
P. M\'esz\'aros, B.-B. Zhang, and P. Veres for discussing
their work with us, and A. Beloborodov, F. Daigne, D. Giannios,
P. Kumar, R. Mochkovitch, A. Pe'er, and F. Ryde for useful discussion.
This work was partially supported by NSF AST-0908362,
National Natural Science Foundation of China (Grants No. 11025313
and 11063001), the ``973" Program of China (2009CB824800), the Guangxi
Natural Science Foundation (2010GXNSFA013112 and 2010GXNSFC013011),
and the special funding for national outstanding young scientist
(Contract No. 2011-135). XFW is supported by One-Hundred-Talent
Program of Chinese Academy of Sciences.

\end{document}